\documentclass[sigconf]{acmart}
\AtBeginDocument{%
  }

\newcommand{\paratitle}[1]{\medbreak\noindent\textbf{#1}}

\copyrightyear{2026}
\acmYear{2026}
\setcopyright{cc}
\setcctype{by}
\acmConference[RecSys '26]{20th ACM Conference on Recommender Systems}{September 27-October 02, 2026}{Minneapolis, MN, USA}
\acmBooktitle{20th ACM Conference on Recommender Systems (RecSys '26), September 27-October 02, 2026, Minneapolis, MN, USA}
\acmDOI{10.1145/3773078.3831734}
\acmISBN{979-8-4007-2284-4/2026/09}

\begin{document}

\title{A Position Paper on Recommender Systems in the Era of Autonomous Agents}

\author{Aixin Sun}
\orcid{0000-0003-0764-4258}
\affiliation{%
  \institution{Nanyang Technological University}
  \city{Singapore}
  \country{Singapore}
}
\email{axsun@ntu.edu.sg}

\begin{abstract}
For decades, recommender systems have been optimized to serve human users. However, the rapid deployment of autonomous agents introduces a paradigm shift: recommendation consumers are predicted to be increasingly a mixture of humans and authorized agents acting on their behalf. This position paper reviews insights from prior human-centric RecSys research and outlines the transition to this hybrid environment. In our discussion, we treat agents as independent, user-aligned assistants that act as end-users of recommender systems, rather than platform-built components. We characterize the resulting tripartite interactions among humans, agents, and platforms, highlighting the dynamics that can arise across these relationships. This shift presents new opportunities for RecSys research, while introducing unique challenges for evaluation, alignment, and system design.
\end{abstract}



\maketitle

\section{From Human-centric to Mixed Interaction}
\begin{figure}
    \centering
    \includegraphics[clip, trim=0.4cm 6cm 16cm 0.1cm, width=0.8\linewidth]{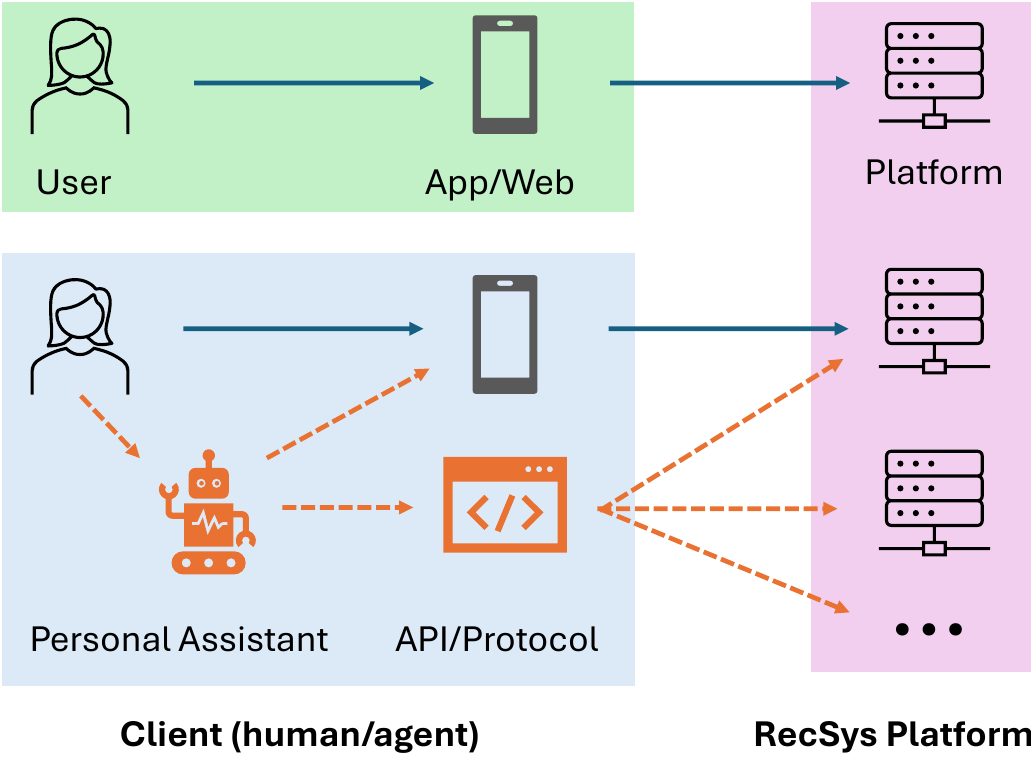}
    \caption{The evolution of RecSys interaction: traditional direct engagement (top) versus the emerging mixed ecosystem where users delegate tasks to autonomous agents, which interact with multiple platforms via GUI or API (bottom).}
    \Description{The evolution of RecSys interaction: traditional direct engagement (top) versus the emerging mixed ecosystem where users delegate tasks to autonomous agents, which interact with multiple platforms via GUI or API (bottom).}
    \label{fig:interaction}
   
\end{figure}

Recommender Systems (RecSys) research is a cornerstone of the SIGCHI community, representing the critical intersection of algorithmic intelligence and human decision-making. Traditionally, the foundational architecture of RecSys has operated on a core premise: the end-user is a human, engaging with a platform through specific interfaces.  The top left-panel in Figure~\ref{fig:interaction} illustrates this user-platform interaction. Engagement was strictly linear and human-driven, with users interacting directly with the recommendation platform through its graphical interface, e.g., web or mobile app.

The bottom-left panel depicts the emerging mixed ecosystem, driven by personal AI assistants such as ChatGPT Agent Mode, Claude Cowork, OpenClaw, and similar alternatives. While direct human-to-platform interaction persists, users can now delegate complex, objective-driven tasks, such as online purchasing, to fully automated personal assistants~\cite{AmineAgentBuying25,AiAgentEcon26,AgenticMarket26,li2026moltbook,Bichler2026}. These autonomous agents exhibit a dual modality of platform interaction. They can visually navigate and execute tasks directly through the web or mobile application interface, mimicking human browsing. Alternatively, when supported by the platform, they can bypass the user interface and communicate directly via dedicated APIs.\footnote{Existing RecSys platforms may be reluctant to provide such API access. However, the growing popularity of agent-centric interaction may create opportunities for entirely new, agent-native business models. A similar transition is already visible in web search, where new services target LLMs rather than human users and adopt completely different business models from traditional search.} As shown by the fan-out in the diagram, this protocol-driven approach empowers the agent to simultaneously query, compare, and aggregate recommendations across multiple platforms, transforming the recommendation landscape into an agent-mediated comparative analysis. This envisioned mixed ecosystem calls for a fundamental re-evaluation of recommender systems: their core roles, objective functions, and optimization strategies.

In this position paper, we define an agent as an autonomous personal assistant that operates on the client side and acts exclusively on behalf of the user. This framing explicitly distinguishes our focus from the prevailing literature, where LLM agents are deployed within the recommendation platform itself to enhance the recommendation system (i.e., agentic recommender systems)~\cite{ZhuLLMAgentSurvey25,huang2025agenticrs,RecommenderAgent25,LLMEmportRec25,TriRec26,huang2025survey,lin2026roadmap}. Rather than competing, we view these two classes of agents as independent and complementary, as they are designed for fundamentally different purposes.

\section{RecSys Tasks and Optimization Objectives}

Before considering the role of agents, we first revisit the problem formulation of a typical RecSys task~\cite{sun2026task}.
\begin{definition}[Recommendation Task]
\label{def:final}
Let $U$ be a set of users and $I$ be a set of items. A recommender system aims to produce a ranked list of items for a user $u$ at time $t$, based on user-item interactions $U \times I$ that are available at $t$, and the conditions specified on the candidate items:
\begin{equation}\label{eqn:selection}
\left\langle u, t, (U\times I)_{\leq t}, s(I_{\leq t}) \right\rangle \rightarrow R_t^u    
\end{equation}
\end{definition}
In Equation~\ref{eqn:selection}, $(U\times I)_{\leq t}$ and $I_{\leq t}$ refer to the interactions and candidate items that are available at timestamp $t$. The selection condition on the candidate items, denoted as $s(\cdot)$, represents selection criteria derived from a user's interaction history, such as repeated consumption, or specified by the user through temporal, spatial, or other attribute-based constraints.  $R_t^u$ represents the ranked items for user $u$ at time $t$.

In the literature, recommender systems are traditionally categorized by their underlying algorithms: content-based, collaborative filtering, and hybrid approaches~\cite{TkdeSurvey05}. While this taxonomy provides a useful high-level overview of technical solutions, it lacks a human-computer interaction (HCI) perspective aligned with the SIGCHI community. Following our recent survey of real-world deployments~\cite{zou2025survey}, we  classify RecSys tasks into two major categories: 
(i)
a \textbf{transaction-oriented recommender system} generates item recommendations with the primary goal of \textit{prompting transactional actions} from users, optimizing for metrics such as conversion, revenue, or purchase likelihood, and (ii) a \textbf{content-oriented recommender system} generates item or content recommendations with the primary goal of \textit{facilitating user consumption and engagement}, optimizing for metrics such as dwell time, clicks, or user satisfaction rather than transactional outcomes. 

Transaction-oriented RecSys aims to facilitate specific actions, such as purchases, bookings, or subscriptions. Its primary goal is to maximize measurable business outcomes like revenue and conversion rates. Content-oriented RecSys, on the other hand, is designed primarily to enhance user engagement and satisfaction by delivering relevant media for \textit{immediate consumption}, such as short-form videos, music streaming, and news. Its core objective is to maximize metrics related to ongoing consumption and retention, including clicks, watch time, and reading duration. While the distinction is not always clear, this classification provides a practical lens for understanding RecSys applications from the user's perspective.

We posit that content-oriented RecSys will primarily continue to serve human users directly. Consumption for entertainment, information gathering, or leisure is inherently a human experience that cannot be meaningfully delegated. While agents may assist with  content tasks such as filtering and scheduling, the value derived from the human experience of engaging with the content itself is not transferable to an agent. This is in contrast to transaction-oriented tasks, where the value lies in the outcome of the transaction rather than the process of reaching it, making delegation not only possible but arguably preferable~\cite{AmineAgentBuying25,PAARS25simiulator,Bichler2026,AiAgentEcon26}. The discussion of agents in this paper focuses specifically on transaction-oriented platforms.

\subsection{When Agents Are in the Picture}
Now, the entire purchasing workflow, from browsing/searching to payment, can be made either by a real user, or by an agent acting on their behalf. For simplicity, in the following discussion, we assume each user will authorize at most one agent to perform transactions across all platforms where they are registered.\footnote{It is technically possible for a user to authorize different agents for different platforms and even more than one agent for the same platform.}

\begin{definition}[Transaction-Oriented Recommendation Task with Agents]
\label{def:transaction_agent}
Let $U$ be a set of human users and $A$ be a set of autonomous agents, where $a_u \in A$ denotes the unique authorized agent acting on behalf of user $u \in U$. A transaction is initiated by a requesting entity $e \in \{u, a_u\}$ at time $t$ with an explicit task goal or prompt, denoted as $g_t$. The recommender system aims to produce a highly optimized response or ranked list $R_t^e$, leveraging the joint historical interactions of both the user and the agent $((U \cup A) \times I)_{\leq t}$, and the task-specific constraints evaluated against the candidate items:
\begin{equation}\label{eqn:transaction_agent}
\left\langle e, t, g_t, ((U \cup A)\times I)_{\leq t}, s(I_{\leq t}, g_t) \right\rangle \rightarrow R_t^e
\end{equation}
\end{definition}

This task introduces the explicit goal parameter $g_t$. When the requesting entity is a human (the case of $e = u$) or a simulated user interacting with the system through the graphical interface, $g_t$ may be latent or loosely defined by human browsing behavior. However, when the requesting entity is an agent (the case of $e = a_u$) interacting with the platform through an API/Protocol, $g_t$ becomes a programmatic constraint (e.g., ``maximize battery life while keeping price under \$1000''). Consequently, the platform's output $R_t^{a_u}$ must transition from a visually persuasive layout (for humans) to a high-density, verifiable data structure (for agents).

In the past, regardless of whether a RecSys task is content- or transaction-oriented, it has been defined from the platform's perspective. That is, the search space consists of currently available items, and the output is what the platform can offer in response to the user's request. Interestingly, as illustrated in Figure~\ref{fig:interaction}, an authorized agent can represent the user to interact with more than one platform. Assume that we have three platforms $\{p_x, p_y, p_z\}$ and all of them are e-commerce services with common items available on them. It is natural that a user and their agent would compare items across platforms. In this sense, the agent now has access to the full picture of the user's past interaction histories with all platforms, rather than the fragmented view obtained by any single platform.\footnote{Because an authorized agent aggregates a user's cross-platform interaction history, it introduces unprecedented privacy and data governance risks. Consequently, this mixed ecosystem necessitates a re-examination of how human values are aligned and embedded within both the delegating agents and the underlying RecSys platforms~\cite{StrayValue2024}. However, such issues are beyond the scope of this paper.} Therefore, the agent has the potential to understand the user's behavior more holistically than any individual platform. 

Formally, user agent $a_u$ constructs a cross-platform candidate set by aggregating candidates received from all platforms $\mathcal{P}$:
\begin{equation}\label{eqn:cross_platform}
\tilde{R}_t = f_{a_u}(\{R_{p,t}\}_{p \in \mathcal{P}}, g_t, h_t^{cross})
\end{equation}
where $f_{a_u}$ represents any aggregation/ranking function operated by the agent, and $h_t^{cross}$ denotes cross-platform behavioral evidence available to the agent (at the user/client side) but not necessarily to any individual platform.

In short, the agent holds a comprehensive, cross-platform view, \textit{i.e.,} accumulating the user's full interaction history, receiving and refining the explicit goal $g_t$, and aggregating responses across platforms. A single platform sees only its own slice: local interaction history with a user (and their agent), its own inventory, and whatever it can infer from the incoming request.

Equations~\ref{eqn:transaction_agent} and~\ref{eqn:cross_platform} characterize a multi-objective ecosystem in which each party pursues a distinct objective. The human seeks to maximize their underlying, often only partially articulated, utility (e.g., a lightweight laptop with long battery life). The agent optimizes a programmatic proxy for that utility (e.g., a laptop weighing under 1.2 kg, offering at least 16 hours of battery life, and belonging to preferred brands A, B, or C). Meanwhile, the platform seeks to maximize business objectives such as conversion and revenue. These potentially competing objectives give rise to three interaction types among the human, the agent, and the platform, which reshape the optimization logic of transaction-oriented RecSys.

\subsection{The Tripartite Dynamics}
\label{sec:tripartite}

Because the agent $a_u$ acts as a centralized aggregator of the user's cross-platform interaction history, the agent possesses a more comprehensive behavioral profile of the user than any individual platform does. The differing objective functions of the human $u$, the agent $a_u$, and the recommendation platform $p$ give rise to three distinct operational frontiers:

\paratitle{Human-to-Platform}.
This remains the traditional domain of RecSys optimization. The platform possesses vast aggregate data and optimizes to maximize transaction volume and revenue. It achieves this by balancing user satisfaction with business objectives, leveraging human cognitive biases, visual hierarchies, and artificial scarcity (e.g., time-limited promotions) to drive more transactions.

\paratitle{Agent-to-Platform}.
When the platform interacts with an agent (the case of $e = a_u$) through API/Protocol, visual persuasion becomes less significant. Because the explicit goal parameter $g_t$ submitted by the agent is highly refined and backed by cross-platform knowledge, platforms will attempt to reverse-engineer the user's global state from $g_t$ to regain their informational advantage. The returned candidate items will compete with those from other platforms, and the agent must verify that all returned items satisfy the specified constraints.  There is also a potential risk that  platforms or third-party sellers may optimize metadata specifically to manipulate agent verification.

\paratitle{Human-to-Agent}.
This introduces a novel intra-user dynamic: the human $u$ must either provide a comprehensive programmatic goal $g_t$ or ensure that their latent desires are accurately mapped to a programmatic goal $g_t$ derived by their delegating agent $a_u$. The primary challenge here is objective misalignment. If a user provides a loosely defined $g_t$ (e.g., ``Buy a reliable laptop for work''), the agent might over-optimize for a proxy metric, such as purchasing the cheapest machine with decent aggregate reviews, while ignoring the user's unstated aesthetic preferences, weight, or brand loyalty. The optimization logic for this interaction requires the agent to actively elicit, clarify, and accurately reflect the human's complete utility function before generating the final ranked list $R_t^{a_u}$ or executing a transaction. Many users may not even know exactly what they  need or prefer, further complicating the agent's task of faithfully representing their true utility.

\section{A Critical Retrospective}
\label{sec:critical}

Before we can effectively conduct research on the mixed human-agent ecosystem, we must critically examine what we have learned from past and current methodologies. Several foundational blind spots in our approach to human-centric RecSys provide critical warnings as we transition to this more complicated setting.

\paratitle{Algorithms Without Systems Thinking}. The RecSys community has overwhelmingly treated recommendation as an isolated machine learning optimization problem, more specifically, as a matrix completion or sequence prediction task, rather than as a holistic HCI challenge~\cite{Zou26hesitation}. This has led to a hyper-focus on algorithmic sophistication at the expense of understanding the ``system'' as a socio-technical artifact. As pointed out by seminal critical reviews, complex neural architectures frequently fail to outperform properly tuned traditional or heuristic baselines \cite{Dacrema2019, Rendle2019,Yambda5b25,McElfreshKV0W22}. By ignoring the human factors, such as  context of consumption and cognitive load, we have built highly sophisticated algorithms that optimize for mathematical proxies rather than genuine user utility.

\paratitle{Static Data in a Dynamic World}. A major driver of this algorithm-centric focus is the field's chronic reliance on a narrow set of outdated offline datasets (e.g., MovieLens, Amazon Product Data). The lack of high-quality, real-world, and dynamic datasets has forced researchers to evaluate modern algorithms on historical snapshots of user behavior that do not reflect contemporary platform dynamics. This reliance creates a self-reinforcing loop: algorithms are designed to win leaderboards on static, outdated datasets, creating an illusion of progress while failing to generalize to dynamic, real-world deployment environments \cite{Jannach26gap,EvaluationSurvey22, zou2025survey}.

\paratitle{What a Dataset Captures vs. What a Task Demands}. Another oversight in past research is the ignorance of what a dataset truly represents. There is frequently a severe mismatch between what a dataset captures and the actual recommendation task being evaluated. Recalling our earlier categorization, researchers may use click-stream data (which is inherently content-oriented and driven by curiosity or visual salience) to evaluate models intended for transaction-oriented tasks (which require high-intent purchasing decisions), because researchers often lack a clear understanding of the differences in RecSys tasks. A dataset only tells us \textit{what} a human user interacted with, completely obfuscating \textit{why} they interacted with it, i.e., the latent goal $g_t$. For instance, although a large number of papers conduct evaluation on the widely used MovieLens dataset, the data collection process may not truly represent real-world RecSys settings~\cite{MovieLensDataset}. Applying static historical interactions to infer future behavior without understanding the differing nature of RecSys tasks leads to optimizing against an incorrect proxy objective.

An example of dataset-task misalignment is the study on Conversational RecSys. Because organic, multi-turn conversational logs from deployed commercial platforms are rarely publicly available, the community has heavily relied on simulated datasets, such as crowdsourced, role-playing dialogues or algorithmic user simulators. However, these simulations may largely fail to reflect reality. These simulations assume a perfectly rational user: one with static preferences who communicates clearly. Real users behave very differently. They change their minds, give ambiguous feedback, or abandon tasks mid-way, all commonly observed in real-world deployments~\cite{zou2025survey}. Optimizing for a constrained, simulated user yields algorithms that perform well on benchmark leaderboards but often fail to generalize to the unpredictability of real-world environments.

\paratitle{The Evaluation Validity Crisis}. This mismatch is compounded by a lack of consistency in how evaluation itself is conducted across the community, leading to difficulty in finding suitable baselines~\cite{RecBaselines23}. Offline metrics (such as Recall@K or NDCG) are routinely used to claim state-of-the-art performance, yet these metrics famously exhibit weak or negative correlation with online A/B testing results and real-world user satisfaction \cite{Jannach2019_business}. When we evaluate systems solely on their ability to predict a missing interaction in a static matrix, we ignore the interactive, multi-turn nature of human decision-making~\cite{sun2026task}. Even within a single evaluation setup, methodological shortcuts further undermine the validity of reported results. For example, many experimental settings suffer from data leakage issues due to ignoring the chronological timeline~\cite{DataLeakage23,Le25Dont,RecNextEval26}. This is also why we explicitly added the timestamp $t$ in our task definitions.

\section{Research Opportunities}
\label{sec:research}

The transition to a mixed human-agent ecosystem demands more than new algorithms: it requires a fundamental restructuring of how RecSys problems are defined, evaluated, and validated. 

The lessons of Section~\ref{sec:critical} are instructive: the field's past failures stemmed not from a lack of sophisticated models, but from misaligned objectives, mismatched datasets, and evaluation frameworks that rewarded leaderboard performance over real-world utility. These are not mistakes to be inherited by the agentic era.

\paratitle{From Preference Modeling to Task Execution}. Historically, RecSys has modeled long- and short-term user preferences to predict the next click, often paying little attention to application- or domain-dependent nuances~\cite{sun2026task}. In the agentic era, this paradigm is insufficient for transaction-oriented systems. When an agent interacts with a platform, it presents an explicit, structured goal $g_t$, not a behavioral trace to be decoded. The research challenge therefore shifts from inferring what the user wants, to parsing and satisfying multi-dimensional, constraint-based task specifications. A concrete open question is: how should a platform's ranking function be redesigned to jointly optimize for constraint satisfiability (verifiable by the agent) and user utility (latent to the platform)? The user's utility should be inferred from not only $g_t$ but also the past traces of the user with the platform. This also demands domain-sensitive system design, since the constraints governing food delivery, hotel booking, and electronics procurement are structurally distinct~\cite{sun2026task}.

\paratitle{From Offline Prediction to Interactive, Protocol-level Evaluation}. The field's reliance on Recall@K and NDCG on static historical logs is already a poor proxy for human utility; it is unsuitable for evaluating agent-to-platform interactions, where correctness is logical rather than probabilistic. We need a new evaluation paradigm that measures the utility of a retrieved batch against the agent's explicit constraints $s(I_{\leq t}, g_t)$, supports multi-turn iterative querying, and can detect adversarial metadata manipulation. Concretely, the community needs benchmark environments in which an agent with a ground-truth utility function issues structured queries to a simulated platform and is evaluated on end-to-end task completion, not item-level prediction accuracy. Such benchmarks must be designed to reflect genuine agentic behavior (e.g., partial specifications, constraint revisions mid-task, and cross-platform arbitrage),  rather than idealized, single-turn interactions.

\paratitle{From Single-platform Retrieval to Cross-platform Brokerage}. Individual platforms have historically defined the search space of recommendation. The agent-side brokerage layer in this new mixed ecosystem opens an entirely new research front: how should an agent aggregate and re-rank across heterogeneous platform responses, each optimized for the platform's own objectives? Key questions include how to detect and correct for platform-side bias in returned candidates, how to weight cross-platform behavioral evidence $h_t^{cross}$ when constructing a unified ranking, and how to design protocols that are machine-readable and resistant to gaming. 

\paratitle{From Task Delegation to Preference Elicitation}. The Human-to-Agent alignment frontier introduces a challenge that no amount of platform-side optimization can solve: if the goal $g_t$ delegated by the human is underspecified, the agent will optimize faithfully for the wrong objective. Research is needed on interactive elicitation mechanisms, in the form of lightweight dialogue protocols, through which the agent proactively surfaces constraint ambiguities, infers latent preferences from past transaction history, and constructs a utility function that the user can inspect and correct before the agent engages external platforms. A concrete research question is: given a sparse history of cross-platform transactions and a loosely specified prompt, what is the minimum elicitation dialogue required to reduce goal misalignment below a measurable threshold? This is closely related to work on Conversational RecSys but differs in a critical way: the goal is not to recommend an item, but to specify a procurement task precisely enough that any competent agent would execute it correctly.

\section{Conclusion}

The rapid emergence of autonomous agents creates a highly plausible near-future setting in which recommender systems must simultaneously serve both human users and software intermediaries. In this position paper, we argued that this shift is not merely a change in the user interface. Rather, it fundamentally restructures the recommendation landscape into a complex, tripartite ecosystem spanning human preferences, agent-driven mediation, and platform optimization. 
The four research directions outlined share a common thread: the move from implicit, behaviorally inferred objectives to explicit, verifiable, constraint-based task execution. 
Advancing these directions will require datasets that accurately represent real-world, agent-mediated recommendation scenarios. In the early stages of this paradigm shift, simulation could be the only viable path to such data. The lessons remain applicable: simulations must reflect genuine agentic behavior, including partial goal specifications, mid-task constraint revisions, and adversarial platform responses, rather than the idealized interactions. 

The community that built RecSys for humans now has both the opportunity and the responsibility to rebuild it for a world in which the consumer is not always human. Issues of privacy, user consent, and cross-platform data governance should also be considered or reconsidered along the way.
\balance

\bibliographystyle{ACM-Reference-Format}
\bibliography{RecSys}

\end{document}